\documentclass[fleqn,usenatbib]{mnras}

\usepackage{newtxtext,newtxmath}
\usepackage[T1]{fontenc}
\usepackage{orcidlink}

\DeclareRobustCommand{\VAN}[3]{#2}
\let\VANthebibliography\thebibliography
\def\thebibliography{\DeclareRobustCommand{\VAN}[3]{##3}\VANthebibliography}
\usepackage{graphicx}	% Including figure files
\usepackage{amsmath}	% Advanced maths commands

\title[Primordial planet spin]{Primordial planet spin driven by boundary layer effects in a decretion disc}

\author[R. G. Martin et al.]{
Rebecca G. Martin,$^{1,2}$\orcidlink{0000-0003-2401-7168}\thanks{E-mail: rebecca.martin@unlv.edu}
Stephen H. Lubow,$^{3}$\orcidlink{0000-0002-4636-7348}
David Vallet,$^{1,2}$\orcidlink{0000-0002-0543-6730}
Madeline Overton,$^{1,2}$\orcidlink{0009-0000-7649-0593}\newauthor
\, Stephen Lepp,$^{1,2}$\orcidlink{0000-0003-2270-1310}
Zhaohuan Zhu$^{1,2}$\orcidlink{0000-0003-3616-6822}
and
Shunquan Huang$^{1,2}$\orcidlink{0000-0002-7276-3694}
\\
$^{1}$Nevada Center for Astrophysics, University of Nevada, Las Vegas,
4505 South Maryland Parkway, Las Vegas, NV 89154, USA\\
$^{2}$Department of Physics and Astronomy, University of Nevada, Las Vegas,
4505 South Maryland Parkway, Las Vegas, NV 89154, USA\\
$^{3}$Space Telescope Science Institute, 3700 San Martin Drive, Baltimore, MD 21218, USA
}

\date{Accepted XXX. Received YYY; in original form ZZZ}

\pubyear{\the\year{}}

\begin{document}
\label{firstpage}
\pagerange{\pageref{firstpage}--\pageref{lastpage}}
\maketitle

% Abstract of the paper
\begin{abstract}
Accretion of material from a protoplanetary disc on to a forming giant planet can spin the planet up to close to its breakup rate, $\Omega_{\rm b}=(G M_{\rm p}/R_{\rm p}^3)$, where $M_{\rm p}$ is the mass and $R_{\rm p}$ is the radius of the planet.   After the protoplanetary disc dissipates, the rapidly rotating planet may eject a decretion (outflowing) disc in a similar way to a Be star. Boundary layer effects in a hydrodynamic disc allow for decretion disc formation at spin rates below the breakup spin rate of the planet. The decretion disc exerts a torque on the planet that slows its spin to an equilibrium value that is sensitive to the planet temperature.  By considering steady state circumplanetary decretion disc solutions, we show that the equilibrium spin rate for planets is around $0.4\,\Omega_{\rm b}$ for $H/R=0.2$ and around $0.2\,\Omega_{\rm b}$ for $H/R=0.3$, where $H$ is the disc scale height at radius $R$. These values are in line with the spins of the giant planets in the solar system and observed exoplanet spins. 
\end{abstract}

% Select between one and six entries from the list of approved keywords.
% Don't make up new ones.
\begin{keywords}
accretion, accretion discs -- planets and satellites: formation -- planets and satellites: gaseous -- planet-disc interactions
\end{keywords}

%%%%%%%%%%%%%%%%%%%%%%%%%%%%%%%%%%%%%%%%%%%%%%%%%%

%%%%%%%%%%%%%%%%% BODY OF PAPER %%%%%%%%%%%%%%%%%%

\section{Introduction}

A giant planet must form while the protoplanetary gas disc is present since the outer layers of the planet are accreted from the disc. Once the planet has mass greater than that of about Neptune, the planet carves out a gap in the proptoplanetary disc \citep{Lin1986,DAngelo2002,Bate2003,Youdin2025}. Accretion of material on to the planet continues through streams and a circumplanetary disc forms around the planet \citep{Artymowicz1996,DAngelo2003,Papaloizou2005,Machida2010,Szulagyi2014}. During this early circumplanetary disc phase, the planet may be spun up to close to its break up spin rate \citep[e.g.][]{Batygin2018,Takaoka2023}.
The breakup angular frequency for a planet is $\Omega_{\rm b}=(GM_{\rm p}/R_{\rm p}^3)^{1/2}$, where $M_{\rm p}$ is the mass of the planet and $R_{\rm p}$ is the radius of the planet.  
However, in the solar system the giant planets have spins that are much below the breakup rate.  Jupiter, Saturn, Neptune and Uranus  have $\Omega_{\rm p}/\Omega_{\rm b}=0.3$, 0.39, 0.17, and 0.16, respectively. Exoplanet oblateness can be detectable from transit constraints \citep[e.g.][]{Seager2002,Barnes2003,Zhu2014oblate,Biersteker2017,Akinsanmi2020,Berardo2022,Liu2025} and measured spin rates have similarly low values to the solar system \citep[e.g.][]{Bryan2018,Bryan2020,Xuan2020}. 

Rapidly rotating stars Be stars may have a similar spin evolution to rapidly rotating planets. Be stars are thought to have been spun up through accretion during the evolution of a companion star \citep[e.g.][]{deMink2013} and are observed to have Keplerian outflowing decretion discs \citep[e.g.][]{Lee1991,Pringle1991,Okazaki2002,Porter2003,Rivinius2013,Carciofi2008}.    This can be a result of boundary layer effects in geometrically thick accretion disc that connects to a rotationally flattened star. The angular frequency of the material in the boundary layer is less than the angular spin frequency of the star. This allows material to viscously flow outwards at the stellar equator into a decretion disc even though the star rotates at less than its breakup rate. This was first demonstrated with 2D hydrodynamical simulations \citep{Hertfelder2017,Dong2021} and more recently \cite{Martin2025} developed a 1D model. Formation of the decretion disc requires a sufficiently high spin rate, high temperature and low levels of infall accretion. The decretion disc exerts a torque on the star that spins it down to an equilibrium value.

This model of the formation of the decretion disc is purely hydrodynamic. However, if a planet has a sufficiently strong magnetic field, the inner parts of a circumplanetary disc may be magnetically truncated.  Magnetospheric accretion occurs when the material flows from the disc to the planet along the field lines \citep{Zhu2015,Batygin2020,Cevallos2025}. If the inner edge of
the disc rotates more slowly than the planet, the circumplanetary
disc torquing by the magnetic field can spin the planet down \citep{Batygin2018,Ginzburg2020}.  Although the magnetic field of Jupiter could be strong enough to truncate the disc, the magnetic fields of Saturn, Uranus, and Neptune were too weak \citep{Stevenson1983,Christensen2010,Wei2024}.

In this Letter, we apply the rapidly rotating star model to a rapidly rotating planet. A planet may eject a decretion circumplanetary disc, after the protoplanetary disc has dissipated and the infall into the Hill sphere has decayed. Spin-out discs around giant planets have been discussed before although, in general, previous models have required the planet to spin at breakup \citep[e.g][]{Korycansky1991,Ward2010}.   In Section~\ref{eq} we calculate the equilibrium spin rate for a planet by considering the torque on the planet from the circumplanetary decretion disc that acts to slow the spin of the planet.
We show that the predictions of the hydrodynamic decretion disc model are in good agreement with the spin rates of the giant planets in the solar system and the observed exoplanet spins.  In Section~\ref{HR} we discuss the expected values of the disc aspect ratio for these circumplanetary decretion discs. Finally in Section~\ref{conc} we draw our conclusions.

\section{Equilibrium spin rate of the planet}
\label{eq}

\begin{figure}
    \centering
    \includegraphics[width=\linewidth]{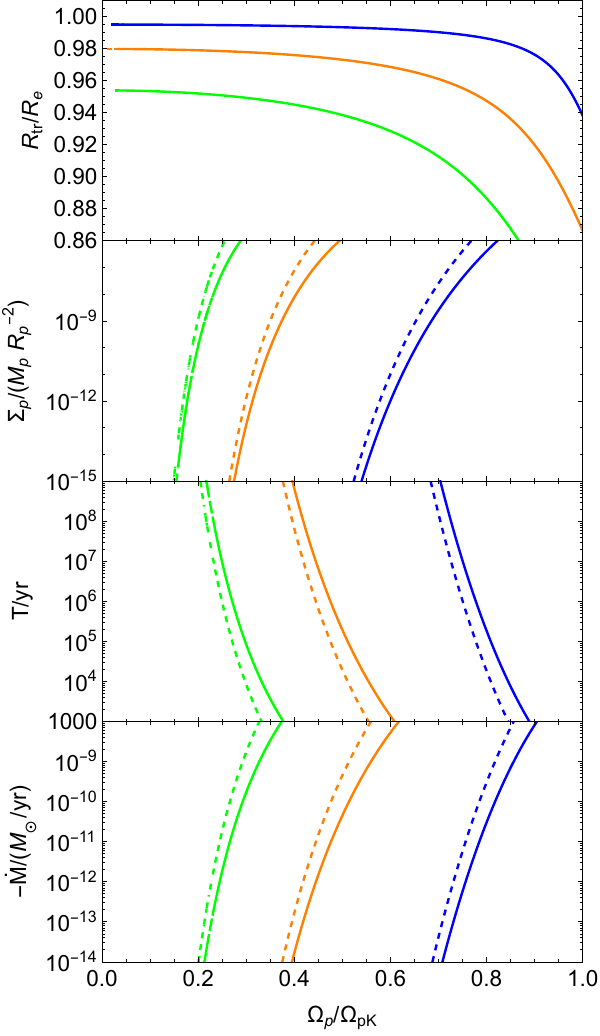}
    \caption{In each panel the blue lines have $H/R=0.1$, the orange lines have $H/R=0.2$ and the green lines have $H/R=0.3$. The solid lines have $X=0.01$ and the dashed lines have $X=0.1$, where $X$ is the ratio of the planet density at $R=0.9\,R_{\rm p}$ to the average planet density. Upper panel: The transition radius from the planet to the disc where the height of the rotationally flattened planet equals the disc scale height $H$ in units of the equatorial radius of an isolated planet, $R_{\rm e}$. Second panel: The surface density of the planet at $R=R_{\rm tr}$. Third panel: The timescale that the planet spin changes with $\alpha=0.01$. Lower panel: The decretion rate for a Jupiter mass planet with $j=17.6$ and $\alpha=0.01$.  }
    \label{fig:combine}
\end{figure}

\begin{figure}
    \centering
    \includegraphics[width=\linewidth]{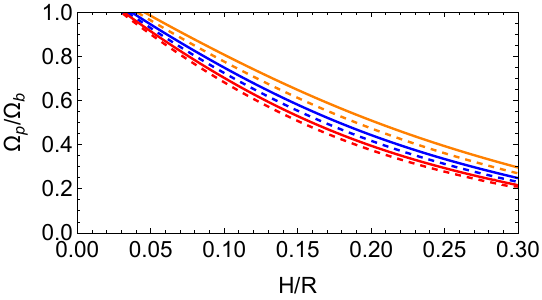}
    \caption{The equilibrium spin rate at $T=10^7\,\rm yr$ for $\alpha=0.01$ (blue) and 0.0001 (orange) and at $T=10^{9}\,\rm yr$ with $\alpha=0.01$ (red lines). }
    \label{fig:omcrit}
\end{figure}

We follow the methods of \cite{Martin2025} to find the equilibrium spin rate for a planet. Motivated by the hydrodynamic simulations of \cite{Dong2021}, \cite{Martin2025} found analytic approximations to the steady state decretion disc flow from a rapidly rotating body. The upper layers of the rotating body rotate with frequency $\Omega_{\rm p}$ and are in hydrostatic equilibrium. The decretion disc that flows from the equator is not assumed to be in Keplerian rotation. The inner boundary condition is that the rotation frequency of the disc, $\Omega$, is equal to the spin frequency of the planet, $\Omega(R_{\rm p})=\Omega_{\rm p}$. In the boundary layer of a decretion disc, material rotates at less than Keplerian frequency and $d\Omega/dR<0$. The material at the equator viscously spreads into a disc, despite the rotation of the central body being less than the break up rate. The decretion disc exerts a torque on the central body that slows the spin of the planet down to an equilibrium value where the torque becomes negligible.

The planet spins rapidly and the flattened planetary surface has a shape defined by
\begin{equation}
    z_0(R)=\left[\left(\frac{1}{R_{\rm polar}}-\frac{\Omega_{\rm p}^2R^2}{2G M_{\rm p}}\right)^{-2}-R^2 \right]^{1/2}
\end{equation}
\citep{Paczynski1991}, where $R_{\rm polar}=z_0(0)$ is the polar radius of the planet. For an isolated planet, the equatorial radius of the planet, $R_{\rm e}$, is defined by $z_0(R_{\rm e})=0$. The radius that marks the transition from the planet to the disc, $R_{\rm tr}$, is where the height of the planet above the equatorial plane is equal to the disc scale height, $H$. The aspect ratio, $H/R$ of a circumplanetary disc may be much larger than the local protoplanetary disc \citep[e.g.][]{MartinandLubow2011}. The transition radius, scaled to the equatorial radius of the planet, is shown in the top panel of Fig.~\ref{fig:combine}. The transition radius is smaller for more rapidly rotating planets and for thicker discs.  

To model the outermost layers of the planet at the equator in $R=0.9\,R_{\rm p}$ to $R=R_{\rm p}$, we solve analytically the equation of hydrostatic equilibrium for a rotating planet in the limit $R \gg z$ assuming a constant sound speed in the layer (equation 24 in \citet{Martin2025}). We define the ratio $X$ to be the density at $R=0.9\,R_{\rm p}$ to the average density of the planet. We consider two values, $X=0.01$ and $X=0.1$.
The surface density of the planet at the transition radius, $\Sigma_{\rm p}$, is shown in the second panel of Fig.~\ref{fig:combine}. This surface density provides a boundary condition for the disc surface density at the inner disc radius $R=R_{\rm p}$. This density is scaled to the mass of the planet and the radius of the planet. 

We model the decretion disc with equations for conservation of angular momentum and a radial equation of motion (see equations 5 and 6 in \cite{Martin2025} and also \cite{Popham1991,Paczynski1991,Lee2013}). For a rapidly rotating planet, an analytic approximation to the steady state decretion disc can be found by assuming that $\Omega \approx \Omega_{\rm p}$ in the boundary layer and then the viscous torque at the inner edge of the disc is calculated \citep[see sections 2.6 and 3.1 in ][]{Martin2025}. The viscosity is assumed to have a constant \cite{SS1973} $\alpha$ parameter throughout the disc. This may be driven by the magneto-rotational instability \citep{BH1991} since $d\Omega/dR<0$ in the boundary layer of a decretion disc \citep[see][]{Martin2025} although hydrodynamic instabilites may also contribute to the viscosity in the boundary layer \citep[e.g.][]{Armitage2002magnetic,Belyaev2012,Belyaev2013,Dittmann2021,Coleman2022,Coleman2022b,Fu2023,Fu2024,Dittmann2024}.

 We find the torque on the planet from a decretion disc that acts to slow the spin of the planet. The third panel of Fig.~\ref{fig:combine} shows the timescale to change the spin of the planet
 \begin{equation}
     T=\frac{J_{\rm p}}{G},
 \end{equation}
where  the angular momentum of the planet is $J_{\rm p}=0.076\, M_{\rm p}R_{\rm p}^2 \Omega_{\rm p}$ and the viscous torque is
\begin{equation}
G=2\pi R \nu \Sigma R^2 \frac{d\Omega}{dR}
\end{equation}
evaluated at the disc inner radius. The viscosity is parameterised with the \cite{SS1973} $\alpha$ parameter as
\begin{equation}
    \nu=\alpha \left(\frac{H}{R}\right)^2 R^2\Omega_{\rm K}
\end{equation}
\citep{Pringle1981}, where the Keplerian angular frequency is $\Omega_{\rm K}=\sqrt{G M_{\rm p}/R^3}$.
 
 If the age of the system is $>T$, then the planet should be spun down to the equivalent spin rate shown in the third panel of Fig.~\ref{fig:combine}.   Note that this of course does require the planet to be spun up to larger values during the protoplanetary disc phase. Therefore, this value can be seen as an upper limit to the spin of a planet. The equilibrium spin of the planet is highly sensitive to the disc aspect ratio. We discuss this further in Section~\ref{HR}.

\begin{figure*}
    \centering
    \includegraphics[width=0.49\linewidth]{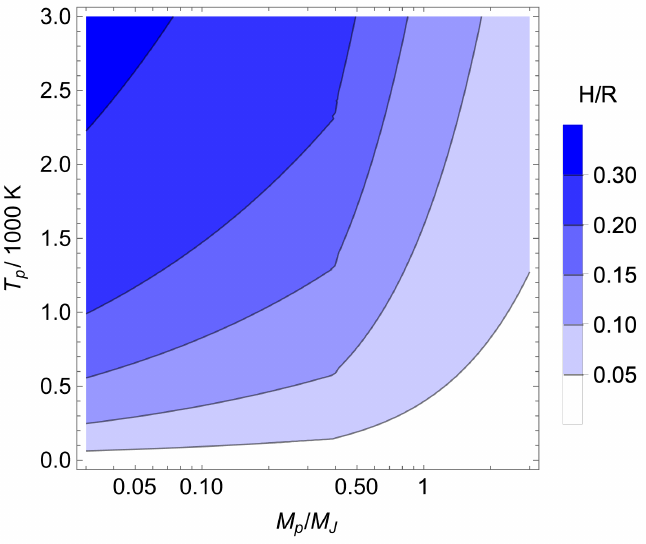}
   \includegraphics[width=0.49\linewidth]{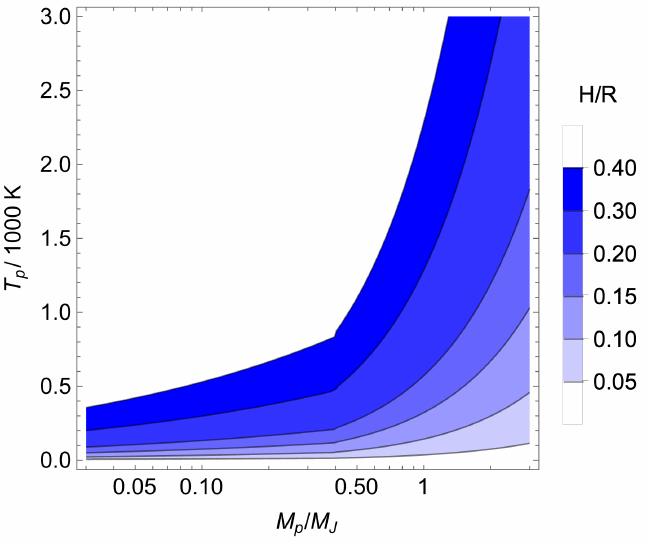}
    \caption{Disc aspect ratio, $H/R$, as a function of the planet mass and planet temperature at $R=R_{\rm tr}$ for $\Omega_{\rm p}=\Omega_{\rm b}$ (left) and $\Omega_{\rm p}=0.3\,\Omega_{\rm b}$ (right).}
    \label{fig:HR}
\end{figure*}

There is a parameter $j$ that appears in the conservation of angular momentum equation. It is the dimensionless rate of flow of angular momentum through the disc divided by the rate of flow of mass through the disc \citep[see e.g.][]{Martin2025}. Its value does not affect the timescale for the spin evolution, but it does affect the decretion rate through the circumplanetary disc.  For a steady state decretion disc solution, material must be removed from the disc at some outer radius, $R_{\rm t}$, and in this case $j\approx (R_{\rm t}/R_{\rm p})^{1/2}$.
 We assume that the outer disc is truncated at a radius of $R_{\rm t}$. For the parameters of Jupiter we find
\begin{equation}
j=
17.6 \, 
\left(\frac{R_{\rm t}}{0.4\, R_{\rm H}} \right)^{1/2}
\left(\frac{a_{\rm p}}{5.2\,\rm au} \right)^{1/2}
\left(\frac{M_{\rm p}}{0.001\,\rm M_\odot} \right)^{1/6}
\left(\frac{M}{\rm M_\odot} \right)^{-1/6},
\end{equation}
where $a_{\rm p}$ is the orbital radius of the planet, and $M$ is the mass of the star.
The Hill radius is
\begin{equation}
R_{\rm H}=a_{\rm p} \left(\frac{M_{\rm p}}{3\,M}\right)^{1/3}.
\end{equation}
The value for $j$ is not very sensitive to the planet mass. For Saturn, Uranus and Jupiter, we find $j=19.5$, $20.2$ and $26.0$, respectively.

The lower panel of Fig.~\ref{fig:combine} shows the decretion rate for a Jupiter mass planet with $j=17.6$ and $\alpha=0.01$. For larger values of $j$, the decretion rate is smaller. The decretion rate becomes negligible as the planet reaches its equilibrium spin rate at large time.  We also note that as the planet cools, the disc aspect ratio decreases. The planet may then have a spin rate below the equilibrium rate at which point no more evolution would be expected through this mechanism.  

Fig.~\ref{fig:omcrit} shows the equilibrium spin rate at $T=10^7\,\rm yr$ (blue and orange lines) and $T=10^{9}\,\rm yr$ (red lines).  The equilibrium spin rate depends sensitively on the disc aspect ratio. The larger the disc aspect ratio, the smaller the equilibrium spin rate. In the next Section we discuss what value should be taken for the disc aspect ratio of a circumplanetary decretion disc and apply this to the planets in the solar system.

\section{Circumplanetary decretion disc aspect ratio}
\label{HR}

 The equilibrium spin rate for the planet is very sensitive to value of the disc aspect ratio that is given by
\begin{equation}
\frac{H}{R}=\frac{c_{\rm s}}{\Omega R},
\end{equation}
where we approximate the angular frequency of the disc as equal to the planet at the disc inner edge, $\Omega\approx \Omega_{\rm p}$. This assumes that the planet is rapidly rotating and the disc is in a steady state. 
The sound speed is
\begin{equation}
    c_{\rm s}=\sqrt{\frac{kT_{\rm p}}{\mu_{\rm m} m_{\rm H}}},
\end{equation}
where $T_{\rm p}$ is the temperature of the surface of the planet, $k$ is the Boltzmann constant, $\mu_{\rm m}$ is the mean molecular weight and $m_{\rm H}$ is the mass of a Hydrogen atom. 

We can express  the  aspect ratio at planet-disc transition radius as
\begin{equation}
\frac{H}{R}=0.244 
\left(\frac{R_{\rm p}}{R_{\rm J}}\right)^{1/2}
\left(\frac{M_{\rm p}}{M_{\rm J}}\right)^{-1/2}
\left(\frac{T_{\rm p}}{1000\,\rm K }\right)^{1/2}
\left(\frac{\Omega_{\rm p}}{0.3\,\Omega_{\rm b}}\right)^{-1},
\label{hr}
\end{equation}
where everything is scaled to the current properties of Jupiter, except the temperature. The temperature should be the temperature at the radius where the disc connects to the star $R=R_{\rm tr}$. This quantity is somewhat difficult to determine since the transition radius itself depends upon the spin rate and the disc aspect ratio. The temperature of the planet may be initially high immediately following its formation.
For example, the temperature of Jupiter may have been around $5000\,\rm K$  and Saturn around $2000-2400\,\rm K$  \citep{Safronov1972}. The temperature may decrease as the planet spins down. We therefore consider a wide range of values for the temperature.

We take the planet mass-radius relationship
\begin{equation}
\frac{R_{\rm p}}{R_{\oplus}}=\begin{cases}
    0.56 \left( \frac{M_{\rm p}}{M_\oplus} \right)^{0.67} & M<127 \, M_\oplus \\
     18.6 \left( \frac{M_{\rm p}}{M_\oplus} \right)^{-0.06} & M>127 \, M_\oplus 
\end{cases}
\end{equation}
\citep{Muller2024}. 
Fig.~\ref{fig:HR} shows the disc aspect ratio at the radius of the planet as a function of planet mass for different planet temperatures. The left panel shows a planet rotating at breakup rate.
Provided that the temperature of the planet at the transition radius is reasonably high ($T_{\rm p} \gtrsim 1000\,\rm K$), the disc aspect ratio of the circumplanetary disc is high, $H/R \gtrsim 0.1$. The right hand panel shows that the disc aspect ratio increases as the spin of the planet decreases to $\Omega_{\rm p}=0.3\,\Omega_{\rm b}$. Therefore a larger disc aspect ratio may be more likely as the planet spins down, if the temperature at the transition radius was fixed.

Considering again Fig.~\ref{fig:omcrit}, a Jupiter mass planet has an equlibrium spin rate of $0.3\,\Omega_{\rm b}$ if $H/R \approx 0.25$.  A lower disc aspect ratio leads to a higher equilibrium spin rate.  Saturn has a higher spin rate, which requires a lower disc aspect ratio, $H/R\approx 0.2$.  The mass of Saturn is about a third of that of Jupiter and the radius is about $0.8$ times that of Jupiter. With equation~(\ref{hr}) we require $T_{\rm p}\approx 400\,\rm K$ for this disc aspect ratio.  On the other hand, Neptune has a radius of $0.34$ times that of Jupiter, a mass of only 0.054 times that of Jupiter. For the spin rate for Neptune (and Uranus) to be this small would require $H/R\approx 0.35$ and $T_{\rm p}\approx 100\,\rm K$. 

\section{Conclusions}
\label{conc}

During the protoplanetary disc phase, a forming giant planet may be spun up to its breakup rotation rate. After the dissipation of the protoplanetary disc, giant planets may eject a decretion disc in a similar way to a rapidly rotating Be star. Boundary layer effects allow the ejection of a decretion disc at spin rates below the breakup rate. The decretion disc provides a torque on the planet that slows the spin to an equilibrium value that is around $0.4\,\Omega_{\rm b}$ if the disc aspect ratio is $H/R=0.2$ and $0.2\,\Omega_{\rm b}$ if the disc aspect ratio is $H/R=0.3$. 
These equilibrium spin rates  are in rough agreement with the spin rates observed in the solar system and in exoplanets. Saturn has a faster spin rate than Jupiter, suggesting that it had a smaller disc aspect ratio which is possible if the temperature of the outer layers of the planet are sufficiently cool. However, magnetic fields may also play a role in the spin-down of Jupiter. Neptune and Uranus have much lower spin rates which may be explained by thicker decretion discs as a result of their much lower planet masses.

The question remains then, if young planets require the formation of a decretion disc in order to spin down, why don't we see the disc in young systems? The timescale for the spin to evolve through this mechanism is very short. On a timescale of tens of thousands of years, the planet can spin down to close to its end value. Therefore we do not expect to observe many planets with high spin rates or large decretion discs. This is different to the case of Be stars that are all observed with a decretion disc. The difference there is that the lifetime of the Be star is short (a few million years) and the decretion rate is much larger.

\section*{Acknowledgements}

We acknowledge support from NASA through grants  80NSSC19K0443 and 80NSSC21K0395.

\section*{DATA AVAILABILITY}
\label{da}
The data underlying this article will be shared on reasonable request to the corresponding author.

%%%%%%%%%%%%%%%%%%%% REFERENCES %%%%%%%%%%%%%%%%%%

% The best way to enter references is to use BibTeX:

\bibliographystyle{mnras}
\bibliography{mnras} % if your bibtex file is called example.bib

% Alternatively you could enter them by hand, like this:
% This method is tedious and prone to error if you have lots of references
%\begin{thebibliography}{99}
%\bibitem[\protect\citeauthoryear{Author}{2012}]{Author2012}
%Author A.~N., 2013, Journal of Improbable Astronomy, 1, 1
%\bibitem[\protect\citeauthoryear{Others}{2013}]{Others2013}
%Others S., 2012, Journal of Interesting Stuff, 17, 198
%\end{thebibliography}

%%%%%%%%%%%%%%%%%%%%%%%%%%%%%%%%%%%%%%%%%%%%%%%%%%

%%%%%%%%%%%%%%%%% APPENDICES %%%%%%%%%%%%%%%%%%%%%

%\appendix

%\section{Some extra material}

%If you want to present additional material which would interrupt the flow of the main paper,
%it can be placed in an Appendix which appears after the list of references.

%%%%%%%%%%%%%%%%%%%%%%%%%%%%%%%%%%%%%%%%%%%%%%%%%%

% Don't change these lines
\bsp	% typesetting comment
\label{lastpage}
\end{document}